\documentstyle[manuscript,aps]{revtex}
\begin{document}

\title{Possible mechanism of the fractional conductance quantization
in a one-dimensional constriction}

\author{V.V. Flambaum \thanks{email address: flambaum@newt.phys.unsw.edu.au} 
and  M.Yu. Kuchiev}

\address{ School of Physics, University of New South Wales,
Sydney 2052, Australia}

%\twocolumn[
%\date{\today}
\maketitle
%\widetext

\tightenlines

%\vspace*{-1.0truecm}

\begin{abstract}
%\begin{center}
%\parbox{14cm}
As it is well known there may arise situations
when an interaction between electrons is attractive. 
A weak attraction should manifest itself
strongly in 1D systems, since 
it can create two-electron bound states.
This paper interprets
the 0.7 $(2e^2/h)$ conductance structure,
observed recently in a one-dimensional 
constriction, as a manifestation of two-electron
bound states  formed in a barrier saddle-point.
The value 
0.75 $(2e^2/h)$  follows naturally 
from the 3:1 triplet-singlet  statistical weight ratio for
the two-electron bound states, if the
triplet energy is lower. Furthermore,
the value 0.75 has to be multiplied by 
the probability T of the bound state formation during adiabatic transmission
of  two electrons into 1D channel ($T\simeq 1$).
If the binding energy is larger than the sub-band energy spacing
 the 0.7 structure
can be seen even when the integer steps
are smeared away by  the temperature. Bound states of several
electrons, if they  exist, may
give different steps at  1/2, 5/16, 3/16  etc in the conductance. 
The latter results are sensitive to the length of 1D system and
 the electron density at the barrier. 
It is not excluded that the fractional conductance
quantization may also appear for a repulsive interaction between electrons. 
In this case the electron level splitting is due to the exchange interaction
with two nearest neighbors in a 1D Wigner crystal.

%\end{center}
\end{abstract}


%] \narrowtext

In a clean one-dimensional (1D)
constriction, where the mean free path is much longer
than the effective channel length, the conductance is quantized in units
of  $2e^2/h$ \cite{Wharam,Wees}. This result can be understood as an
adiabatic electron transmission in a 2D system which is electrostatically
squeezed by a  negative gate voltage into a 1D channel. 
In the papers \cite{Thomas96,Thomas} it has been found that in addition
to the usual quantized conductance plateaus, there is also a structure
at 0.7 $(2e^2/h)$. The aim of the present letter is to present
a  possible explanation for this structure.

The simplest theoretical description of 
the electron behavior in the 1D constriction 
usually starts from the electron {\em gas}  approximation, 
which   neglects interaction between electrons.
Since the transverse degrees of freedom in the 1D constriction 
are quantized, there arises additional simplification
based on the {\em adiabatic} approximation, which neglects
transitions between the transverse degrees of freedom.
This simplest, naive picture describes the constriction 
as a one-dimensional
system of independent electrons in which the electron
transverse degrees of freedom
manifest themselves as separated
{\em transition channels}, see Refs.\cite{Imry}.
In spite of its very simple nature,  this physical picture 
correctly describes
a vital, observed experimentally, see Refs.\cite{Wharam,Wees} 
property of conductivity,
namely the jumps of conductivity with variation 
of the gate voltage. These jumps may be interpreted
as the crossing of the Fermi level of the electron gas with
different transverse energy levels. A crossing provides 
a possibility for the electron gas 
to occupy a new transition channel.
Each channel gives a  contribution to
the conductivity equal to 2$e^2/h$, 
in agreement with results of \cite{Wharam,Wees}.

This simple physical picture has recently  faced a challenge.
Refs.\cite{Thomas96,Thomas} demonstrate that
there exist conductivity jumps which deviate 
significantly  from the naive value
2$e^2/h$. Several different jumps have been reported;
one of the most pronounced among them is the jump $0.7 \times (2 e^2/h$),
the so called 0.7 jump. These new jumps inspire one
to abandon either the electron gas or adiabatic 
approximations, or both of them. 
In this paper we concentrate our attention on a possible violation of the
single-electron approximation.
If electron correlations come into play, 
then, generally speaking, one could have anticipated a dramatic 
increase of the difficulty of the problem and, accordingly,
much more sophisticated explanation of the physical events 
which take place in the constriction. 
 The results of Refs.\cite{Fisher,Ogata}, which uses the ideas of the Luttenger
1D  liquid, can illustrate this point. The main result of this letter is
a statement that  even when  correlations are
strong, there still remains a very 
simple and clear way to explain the origin of the $0.7(2e^2/h)$ jump,
and possibly other noninteger jumps.

The first question to be addressed is the
nature of an interaction between electrons. Obviously there
is  their Coulomb repulsion, but alongside with it 
there may exist also an attraction
caused, for example, by the phonon exchange. 
The total interaction arises due to interplay of
the repulsion and attraction, and can have either sign.
In this letter we consider both possibilities, 
paying more attention to the case of an attraction, and briefly
discussing the repulsive case at the end of the letter.

If one assumes that in the 1D constriction the attraction dominates, then
it should have dramatic consequences, because in a one-dimensional
system any, however weak,
attraction results in creation of a two-electron bound state.
If such pairing is possible, 
then the energy of a 1D bound state should depend
on the spin of the pair, since the Coulomb exchange interaction 
should play a very  essential role, as elaborated below. 
As a result there arises  an energy splitting  between 
the singlet and the triplet bound states 
(if the splitting is strong enough, then the upper bound state can even disappear).
Returning to the transmission channels picture, let us
suppose that the gate voltage is such  that
a single electron has not enough energy 
to occupy some transition channel and 
to overcome the barrier. In the single-electron 
picture this channel does not contribute to
the conductivity.
An existence of the bound states provides a new possibility.
Two electrons can comprise a pair and their binding  energy 
can be used to overcome the barrier.
In other words, a formation of the  
electron pair reduces the effective height of the barrier 
and {\em opens} a channel for a pair.
Moreover, one can   argue that
the corresponding jump should be less than
the single-electron value $2e^2/h$. 
In order to see this suppose that, for example,
the triplet bound state S=1 of a pair
is lower than the singlet S=0 one.
(This assumption agrees with the fact
the triplet bound state
S=1 is favoured by the Coulomb exchange interaction.)
Then we can consider the gate voltage such that
the binding of the triplet bound state provides enough energy to 
overcome the barrier, while the singlet binding
is insufficient.
Then only the triplet pairs give contribution to the
conductance. 
The corresponding jump of 
the conductivity should be $3/4\times(2 e^2/h)$, simply 
because    formation of a triplet bound state statistically
is three times more probable, than formation
of a singlet state.
The value 3/4 is close to 0.7, and there is a way to
reduce it slightly, as discussed below.
We conclude that the 0.7 jump may be explained 
by an existence of a triplet pairing at the top of the barrier.
Additionally, there exists a possibility to have a 1/4 jump. It arises
if the singlet state of the pair is the ground state.

In order to elaborate these arguments let us first
note that
in 1D case with slow electrons the exchange
 Coulomb interaction is practically not suppressed in comparison with
the direct Coulomb interaction, and therefore
the Coulomb repulsion is partly compensated for 
by the exchange interaction.
Indeed, in the 1D case direct ($Q_d$)
and exchange($Q_{ex}$) Coulomb integrals
\begin{equation}\label{dir}
Q_d \sim \int \frac{dz_1 dz_2}{|z_1 - z_2|} 
\end{equation}
\begin{equation}\label{exch}
Q_{ex} \sim \int \frac{\exp{[i(k_1-k_2)(z_1-z_2)]}dz_1 dz_2}{|z_1 - z_2|}  
\end{equation}
have only some weak logarithmic dependence on the parameters
of the problem (transverse localization radius, electron wave vectors $k$ , etc).
Therefore, the ratio $ Q_{ex}/Q_d$
should not differ significantly from 1, and the exchange interaction can
significantly reduce the repulsion.
In this situation
any attraction, for example the one which is induced by the phonon exchange, 
have  better chances to become dominant. Notice that
the above argument is valid for slow
electrons, and therefore becomes more important at the top
of the barrier. If electrons are fast, then the oscillating exponent
in the integrand 
in the right-hand side of Eq.(\ref{exch}) 
diminishes the value of the integral thus reducing
the exchange interaction.
These qualitative arguments support the idea that
an attraction at the top of the barrier is 
a possible option. We will  content with that,
since quantitative development in this  direction needs
more effort than this letter can afford.

In order to calculate
the electron transmission let us start from
the most simple picture of 
the electron gas confined by a one-dimensional smooth semiclassical
 potential.
Suppose that the highest energy level of confined electrons is at the top of
the barrier. Let us consider what happens if an additional longitudinal
 electric
field is applied. This field creates the states above the barrier.
 These above-barrier
states give contribution to the current. In order to calculate this current
one needs to multiply the charge, velocity and density of electrons.
 It is convenient
to fulfill calculations at the top of the barrier, which is supposed to be
 rather
flat. One finds
\begin{equation}
\label{I}
I=e\int \, v_{z}\, dN_{z}=e\int \, v_{z}\frac{2\, dp_{z}}{2\pi \hbar }=
\frac{2e}{h}\int d\epsilon =\frac{{2e}}{h}\epsilon =\frac{2e^{2}}{h}V
\end{equation}
Here \( 2dp_{z}/(2\pi \hbar ) \) is the density of electron states, where
the coefficient 2 comes from two projections of spin. The maximal energy excess
over the barrier \( \epsilon  \) is related to the applied voltage \( V \),
\( \epsilon =eV \). From Eq.(\ref{I}) one derives the conductivity
\begin{equation}
\label{G}
G =\frac{2e^{2}}{h}
\end{equation}
. This is a particular
case of the Landauer formula for the conductivity (see e.g. \cite{Imry}
and references therein)
\begin{equation}
\label{GL}
G =\frac{2e^{2}}{h}\sum_{ij}T_{ij}
\end{equation}
 corresponding 
to the transmission coefficients $T_{ij}=\delta_{ij}$ and one open channel.
The summation here is carried out over \( n \)  open channels.

Let us calculate now the conductivity for the case when there are bound
states of the electron pairs in the channel. 
It is convenient to consider a gas of electrons
in a 1D channel  as a set of 'pairs', each 'pair'
being comprised of two neighbor electrons.
Obviously, thus defined 'pair' for noninteracting electrons 
possesses zero binding energy.
The conductivity Eq.(\ref{I}) can be expressed
in terms of these 'pairs'.
Separating contributions of the singlet
and triplet 'pairs' in the total current $I$ one rewrites
Eq.(\ref{I}) as 
\begin{equation}
\label{IP}
I =\frac{2e^{2}}{h}V \left(\frac{1}{4}T_s +\frac{3}{4}T_t\right)
\end{equation}
In the electron gas approximation 
the singlet ($T_s$) and triplet ($T_t$) transmission coefficients
are both equal to unity, and  Eq.(\ref{IP}) 
coincides with  Eq.(\ref{I}). 
Let us now take into account the interaction between electrons,
supposing that it has an attractive nature.
This interaction produces real pairs 
with some positive binding energy.
The pairs should have a pronounced
singlet-triplet splitting,
since the exchange Coulomb interaction 
is strong, as is discussed above.
The energy splitting has a dramatic influence on the 
transmission coefficients $T_t,T_s$. 
Suppose that the triplet state is lower.
Then the  gate voltage may be set 
is such a way that the triplet state is above
the barrier, while  the singlet state is below the barrier.
This makes the triplet transmission coefficient be 
equal to unity, $T_t=1$,
and the singlet coefficient be negligible, $T_s=0$.
For this case Eq.(\ref{IP}) predicts that
the contribution of the transmission coefficient to
the conductivity is $0.75 (2e^2/h)$.
Note that a more accurate calculation,
which remains outside the scope of this letter,
should give the result $T_t \le 1$.
Indeed,  pair formation requires adiabatic transformation of
the continuum state into the bound state.
Opening of a new channel in the 
single-electron approximation
corresponds to a zero electron
velocity at the top of the energy barrier in a saddle point. 
A purely classical consideration
gives in this case an infinite time for the electron transmission 
through the constriction. In quantum mechanics the transmission 
time, defined as a derivative of the 
transmission phase over the electron energy, remains  infinite as well.
This allows one to speak about the  adiabatic transformation of
the continuum state into the bound state in the case of the smooth barrier.
 For zero-energy electrons
the adiabatic transformation of the inter-electron potential leads to 
 binding with
a probability  1. (One can compare this process with a capture
to the E=0 bound state from the continuum in the two-body 
scattering problem, where 
the cross-section is infinite for zero-energy
particles). For electrons on the Fermi surface the
binding can happen near a single-particle turning point
 in a 1D effective potential and
has the probability that may be slightly smaller than 1. Unbound 
electrons are reflected back thus making the transmission coefficient $T_t$
smaller than 1.

It is instructive to present a simple estimate for the pair energy
as a function of the sizes of the perpendicular directions $L_x$
and $L_y$. Consider, for example, a short -range attractive interaction
 between the particles
that have  potential depth $U_0$ and  range $a$.
The ground state energy of two interacting particles can be estimated
as
\begin{equation}
\label{EL}
E(L)=\frac{\pi^2 \hbar^2}{2m}
\left (\frac{1}{L_x^2}+\frac{1}{L_y^2}+\frac{c}{L^2} \right)
-\frac{U_0 a^3}{L_xL_yL}
\end{equation}
where $c \sim 1$. The length of the bound state $L=L_z$ can be found
from the condition of the minimal energy $\frac{dE}{dL}=0$. The substitution
of this $L$ into Eq. (\ref{EL}) gives:
\begin{equation}
\label{E}
E = \frac{\pi^2 \hbar^2}{2m} \left (\frac{1}{L_x^2}+\frac{1}{L_y^2} \right )
-\frac{2m}{\pi^2 \hbar^2c} \left (\frac{U_0 a^3}{L_xL_y}\right )^2
\end{equation}
In this expression one of the dimensions, say  $L_y$, may be 
taken constant
and another dimension $L_x$ considered as  a function of z. We see that
 both the positive
kinetic energy term and negative attractive interaction create the 
 effective potential $U(z)=E$ for the pair which is
inversely proportional to $L_x^2$.
The attractive
interaction  reduces the height of the effective potential.

  The binding energy of the pair can be larger than 
 the sub-band energy spacing $\sim \pi^2 \hbar^2 / m L_x^2 $.
 This may explain why the 0.7 structure is seen even at the temperatures
when the integer steps in the conductance are smeared away.

Our discussion so far focused on the contribution of
the electron pairs into the conductivity.
However, if there is attraction in 1D
electron system and the length of 1D segment is long enough ,
 then one can anticipate that
alongside with pairs there should arise
bound states with several electrons
\( N \). If we suppose that the many-electron bound state
can exist at the top of the barrier ,
then its contribution to the conductivity 
can be found using the arguments similar to those which 
were lead to Eq.(\ref{IP}). The result reads 
\begin{equation}
\label{gen}
\sigma =\frac{\sum _{S,i}(2S+1)T_{S,i}}{2^N}
\, \left( \frac{2e^2}{h} \right)~,
\end{equation}
where $S$ is a possible spin of the $N$-electron bound state,
$S \le N/2$, $T_{S,i}$ is the transmission coefficient for 
the state with the spin $S$ and $i$ is the index that numerates
 different states with a given $S$;
  the number of different
levels $n(S)$ with a given $S$  is equal to \cite{LL}
\begin{equation}\label{nS}
n(S)= \frac{N ! \, (2S+1)}{\left( \frac{N}{2}+S+1\right) ! 
\left( \frac{N}{2}-S\right) !} ~.
\end{equation}
The factor $2^N$ in the denominator of Eq.(\ref{gen})
is equal to  the total number of 
of all posible states in the system of $N$ electrons,
$2^N = \sum_S (2S+1) n(S)$. 
According to Eq.(\ref{gen}) a bound state with spin $S$
gives a jump of conductivity 
\begin{equation}\label{genj}
\delta \sigma =\frac{(2S+1)}{2^N} ~T_{S,i}
\, \left( \frac{2e^2}{h} \right)~,
\end{equation}
As was mentioned above, the exchange Coulomb interaction 
favours the state with the maximal spin $S = N/2$. 
If we assume that this state is the lowest one, then
we obtain from Eq.(\ref{genj}) that the first
step in the conductivity is equal to 
\begin{equation}\label{genjm}
\delta \sigma =\frac{N+1}{2^N} \, \left( \frac{2e^2}{h} \right)~,
\end{equation}
where for simplicity the transmission coefficient is omitted, since it
is supposed to be close  to unity, $T_{N/2} \simeq 1$.
For the triplet two-electron bound state  $N=2$   
Eq.(\ref{genjm})  reproduces the result
$\delta\sigma= 3/4 \, (2e^2/h)$ in accord with Eq.(\ref{IP});
if more electrons comprise a bound state,
$N=3,\,4,\, 5$, then Eq.(\ref{genjm})
results in $ \delta\sigma = 1/2, \, 5/16, \, 3/16~ (2e^2/h)$ respectively.
If there are bound states with both maximal spin and smaller
spins one can observe several fractional steps. For example,
 the three-electron problem has three eigenstates:
spin 3/2 (statistical weight  $2S+1=4$), and two states with S=1/2 
(statistical weight 2 for each state). Therefore, in this case
 there may be steps for 1/2
and 3/4  $(2e^2/h)$. The first step corresponds to the bound state with
the  total electron
 spin 3/2 and the second step to one of the two states with spin
 1/2 .

 Note that the existence of the multielectron bound states
may depend on the length of the 1D segment. Therefore,
the steps in a 1D wire may be different from the steps in a quantum
point contact.

Now consider the transmission in a magnetic field.
For a weak magnetic field a triplet pair
should exhibit a splitting 
to three levels with magnetic g-factor $g=2g_1$ where $g_1$ is the
single-electron 
g-factor in the materials under study. 
Notice that such splitting has not been discovered so far. 
It is possible, however, 
that the weak-field splitting remains hidden due to finite
widths of the peaks in the transconductance $dG/dV_g$,
where $V_g$ is the gate voltage, reported in Ref.\cite{Thomas}. 
For strong 
magnetic fields the triplet pair can exist  only
for the maximal single-electron spin projections $s_z=1/2$.
For other spin projections the energy of the pair
is insufficient to allow penetration over the top
of the barrier. Moreover, the pair does not form
since the electron with   $s_z=-1/2$ can not reach
the top of the barrier.
In this case
one has the single-electron results (1/2 splitting of the step) with
an additional small feature near 1/4 due to the pair formation in the channel
$S_z=1$.

We have discussed so far a case of attractive interaction between electrons
at the top of the  barrier. 
Let us discuss briefly the case of repulsion.
It is demonstrated in the Ref. \cite{Glazman}
that a 1D Wigner crystal is formed for electron density
smaller than the (Bohr radius)$^{-1}$.
The interaction of an electron with two neigbors
in a segment of unpolarized Wigner crystal
gives a picture which is somewhat similar to that
for a three-electron bound state. In half of the cases
the maximal spin S=3/2 state is formed. Another half corresponds
 to the two spin 1/2 states. To find the energies of these states
let us introduce the Hamiltonian for the interaction of the middle
electron 2 with it's neighbors on both sides 1 and 3.
\begin{equation}
\label{J}
J_{12} {\bf s_1 s_2} + J_{23} {\bf s_2 s_3} 
\end{equation}
This Hamiltonian gives the following energies for the S=3/2 and
two S=1/2 states correspondingly:  
\begin{equation}
\label{EJ3}
E(3/2)= \frac{J_{12}+ J_{23}}{4}
\end{equation}
\begin{equation}
\label{EJ1}
E(1/2)= -\frac{J_{12}+ J_{23}}{4} \pm \sqrt{J_{12}^2+J_{23}^2 -J_{12}J_{23}}
\end{equation}
The ferromagnetic-type exchange interaction corresponds to the negative
$J$ and S=3/2 ground state. When the electron  ``train''
 (a segment of the 1D Wigner crystal) is climbing
 the barrier the distance between the electrons 
is changing. This gives a possibility of $J_{23} \gg J_{12}$. In this case
we have effectively the two-electron picture and the statistical
weight ratio 3:1.
( the splitting between the 3/2 level and one of the 1/2 levels
is    $\sim J_{12}$ and small).  If we consider a long 1D wire
the distance between the electrons is constant and  $J_{23}=J_{12}=J$.
In this case we have three different energy levels ($J/2, 0, -J$) with
 the ratio of the statistical
weights  2:1:1. One may consider this as a hint for the possible
steps at 1/2 and 3/4  $(2e^2/h)$. However, the conductivity
for  the segment of the 1D Wigner crystal may be very different
from that for the free electrons, and such conclusion is at least
 not obvious.
  
 We stress that this simple consideration does not {\em prove} 
that the two-electron and multi-electron effects 
exist. It only shows that they are conceivable. The situation would be much
 simpler if there is an 
electron  trapped by an impurity close to the saddle point.
The exchange interaction beteen this electron and free electron 
provides 3:1 splitting. However,
the study in the Ref. \cite{Thomas} has not revealed any dependence
on the position of the transmission channel. This seems to exclude
any explanation based on the impurity location.
 
   Recently it has been found in   Ref. \cite{Reilly}
that in a true 1D system such as the 2${\mu}$m wire, the conductance
 structure occurs at  0.5 $(2e^2/h)$. There have already been attempts to
derive  this result using the exchange interaction.
 The long wire can be occupied by many
electrons. Recent calculations \cite{Gold,Wang} have shown that there is
a spontaneous spin polarization in a 1D electron gas.
 The energy splitting due to
 spin flip of one electron in the mean
 field of  other electrons can give this structure in the transmission
 coefficient.

This work was supported by the Australian Research Council.
 We are grateful to B. Altshuler, R. Clark, R. Newbury, D. Neilson and 
O. Sushkov for valuable discussions and information about their
results prior to publication.
VVF is grateful to the Special Research Center for Subatomic Structure
 of Matter,
University of Adelaide where part of this work was done.

\end{document}